\documentclass[11pt]{article}
\usepackage{amsmath,amssymb,graphicx}
\usepackage{comment}

\newcommand{\ba}{\begin{alignat}{3}}

\newcommand{\pa}{\partial}

\begin{document}

\begin{titlepage}
\begin{flushright}
\end{flushright}
\begin{center}
  \vspace{3cm}
  {\bf \Large Quantum Aspects of Black Objects in String Theory}
  \\  \vspace{2cm}
  Yoshifumi Hyakutake
   \\ \vspace{1cm}
   {\it College of Science, Ibaraki University \\
   Bunkyo 2-1-1, Mito, Ibaraki 310-8512, Japan}
\end{center}

\vspace{2cm}
\begin{abstract}
One of important directions in superstring theory is to reveal the quantum nature of black hole.
In this paper we embed Schwarzschild black hole into superstring theory or M-theory, 
which we call a smeared black hole, and resolve quantum corrections to it. 
Furthermore we boost the smeared black hole along the 11th direction and construct
a smeared quantum black 0-brane in 10 dimensions.
Quantum aspects of the thermodynamic for these black objects are investigated in detail.
We also discuss radiations of a string and a D0-brane from the smeared quantum black 0-brane.
\end{abstract}

\end{titlepage}

\setlength{\baselineskip}{0.65cm}


\section{Introduction}


Superstring theory is a promising candidate for a theory of quantum gravity\cite{Green:1987mn,Polchinski:1998rr}.
Fundamental objects in the superstring are strings and D-branes, which are approximated by black objects 
in the low energy limit. 
Therefore studies of quantum corrections to strings and D-branes are quite important
to reveal quantum aspects of the black objects.

Quantum corrections in the superstring theory have been investigated from the mid of 
1980s\cite{Gross:1986iv,Gross:1986mw,Grisaru:1986px,Grisaru:1986vi}.
In the low energy or classical limit, interactions of string scattering amplitudes 
are well described by 10 dimensional supergravity.
Once we take into account oscillations or loop effects of strings, however,
the low energy effective theory of the superstring theory should be modified. 
For instance, there are famous $R^4$ terms in type IIA superstring theory\cite{Gross:1986iv}
and Gauss-Bonnet term in type I and heterotic superstring theory\cite{Gross:1986mw}.
The existence of these terms can also be confirmed by calculating the vanishments of $\beta$-functions 
in non-linear sigma model\cite{Grisaru:1986px,Grisaru:1986vi}.
Since M-theory is related to the type IIA superstring theory via circle compactification,
it is expected that there exist $R^4$ terms in the M-theory\cite{Tseytlin:2000sf}.
In fact the existence of $R^4$ terms in the M-theory is confirmed from the viewpoint of the supersymmetry 
in refs.~\cite{deRoo:1992sm}-\cite{Hyakutake:2007sm}.
Applications of these higher derivative terms can be found, for example, in the context of black hole physics\cite{Callan:1988hs}, 
gauge/gravity duality\cite{Gubser:1998nz,Tseytlin:2000sf}, flux compactification\cite{Becker:2001pm} and string cosmology\cite{Maeda:2004vm}.

Recently a progress on the quantum corrections in the type IIA superstring theory has been made 
for D0-brane physics\cite{Hyakutake:2013vwa,Hyakutake:2014maa,Hyakutake:2015rqa}.
In the classical limit, the D0-branes are well described by a black 0-brane solution in type IIA supergravity.
In 11 dimensional supergravity, it can be identified with a M-wave solution which carries Kaluza-Klein momentum along the 11th direction.
Since the M-wave is purely geometrical object, it is possible to examine leading quantum corrections to it 
only by taking into account $R^4$ terms in the M-theory.
In ref.~\cite{Hyakutake:2013vwa}, the quantum corrections to the near horizon geometry of 
the black 0-brane and the M-wave are investigated and quantum aspects of the thermodynamics are revealed.
The asymptotically flat quantum D0-brane is constructed in ref.~\cite{Hyakutake:2014maa}, and it is reconstructed by using
the duality between the type IIA superstring theory and the M-theory in ref.~\cite{Hyakutake:2015rqa}.

In this paper, we study quantum aspects of black holes in 4 dimensions. 
In order to execute these tasks, first we uplift
Schwarzschild black hole solution into 11 dimensions, which is smeared over 7 spatial directions.
Since the smeared black hole is purely geometrical object in 11 dimensions, it is possible 
to examine leading quantum corrections to it and construct quantum black hole solution.
The mass and the entropy of the quantum black hole are determined up to the leading quantum corrections, 
and it is shown that the thermodynamic properties, such as the specific heat, are modified by the quantum effect. 
We execute similar analyses for a smeared quantum black 0-brane and its near horizon geometry. 
Furthermore, Hawking radiations of a string and a D0-brane out of the smeared quantum black 0-brane 
are estimated by calculating the tunnelling probability through the event horizon.

Organization of this paper is as follows.
In section \ref{sec:QBO}, we construct the smeared quantum black hole solution. 
By boosting it along the 11th direction we construct the smeared quantum M-wave solution in 11 dimensions,
and by reducing the 11th direction we obtain the smeared quantum black 0-brane solution in 10 dimensions.
In section \ref{sec:T-QBO}, we discuss the thermodynamics of the smeared black objects.
The Hawking radiations of the string and the D0-brane out of the smeared quantum black 0-brane are discussed in section \ref{sec:rad}.
Section \ref{sec:con} is devoted to conclusion and discussion.
In appendix \ref{sec:solfg}, we give detailed calculations omitted in section \ref{sec:QBO}.
In appendix \ref{sec:radp}, the radiation process of a D$p$-brane from classical black $p$-brane is explained.


\section{Smeared Quantum Black Objects} \label{sec:QBO}


It is important to understand quantum nature of black hole in 4 dimensions.
Since superstring theory or M-theory are formulated as a quantum theory of gravity, 
we consider this problem by embedding the black hole into 10 or 11 dimensions.
In this section, we embed Schwarzschild black hole into 11 dimensions, which is smeared over 7 spatial directions.
We call it a smeared black hole, and take into account quantum corrections originated from the string theory.
We also boost the smeared quantum black hole along 11th direction and construct a smeared quantum M-wave solution.
When the 11th direction is reduced, this solution corresponds to the smeared black 0-brane in 10 dimensions.

It is well-known that the supergravity is realized as the low energy limit of the superstring theory,
and string oscillations and string loop effects give $\ell_s$ (string length) and $g_s$ (string coupling) 
corrections respectively.
In general it is quite difficult to take into account such corrections.
For the black 0-brane, however, we can explicitly construct the quantum solution which incorporate 
1-loop scattering amplitudes of type IIA superstring theory\cite{Hyakutake:2013vwa,Hyakutake:2014maa,Hyakutake:2015rqa}.

Since the black 0-brane can be expressed by the M-wave solution in 11 dimensions, 
which is purely geometrical object and easy to deal with,
we will mainly consider quantum corrections in M-theory.
The low energy limit of the M-theory is approximated by 11 dimensional supergravity\cite{Cremmer:1978km},
and it is known that the leading quantum corrections contain so called $R^4$ terms\cite{Tseytlin:2000sf,Becker:2001pm}.
The bosonic part of the M-theory effective action which is relevant to the geometry is given by
\begin{alignat}{3}
  S_{11} &= \frac{1}{2 \kappa_{11}^2} \int d^{11}x \; e \Big\{ R +
  \gamma \Big(t_8 t_8 R^4 - \frac{1}{4!} \epsilon_{11} \epsilon_{11} R^4 \Big) \Big\} \notag
  \\
  &= \frac{1}{2 \kappa_{11}^2} \int d^{11}x \; e \Big\{ R +
  24 \gamma \big( R_{abcd} R_{abcd} R_{efgh} R_{efgh}
  - 64 R_{abcd} R_{aefg} R_{bcdh} R_{efgh} \notag
  \\
  &\qquad\qquad\qquad\qquad\quad
  + 2 R_{abcd} R_{abef} R_{cdgh} R_{efgh}
  + 16 R_{acbd} R_{aebf} R_{cgdh} R_{egfh} \label{eq:R4}
  \\
  &\qquad\qquad\qquad\qquad\quad
  - 16 R_{abcd} R_{aefg} R_{befh} R_{cdgh}
  - 16 R_{abcd} R_{aefg} R_{bfeh} R_{cdgh} \big) \Big\}, \notag
\end{alignat}
where $a,b,c,\cdots = 0,1,\cdots,10$ are local Lorentz indices.
Note that indices are lowered for simplicity, but of course they should be contracted by the flat metric.
It is confirmed that the above M-theory effective action is surely invariant under the local supersymmetry 
if we add appropriate fermionic terms\cite{Peeters:2000qj}-\cite{Hyakutake:2007sm}.
The gravitational constant and the expansion parameter can be expressed 
in terms of the 11 dimensional Planck length $\ell_p = g_s^{1/3} \ell_s$ as
\begin{alignat}{3}
  2\kappa_{11}^2 = (2\pi)^8 \ell_p^9, \qquad \gamma = \frac{\pi^2\ell_p^6}{2^{11} 3^2}. \label{eq:gamma} 
\end{alignat}
Since the parameter $\gamma$ is written like $\gamma \sim g_s^2 \ell_s^6$, if we reduce the action (\ref{eq:R4}) 
into 10 dimensions, it corresponds to 1-loop effective action for the type IIA superstring theory.
The effective action (\ref{eq:R4}) is reliable when $\gamma$ is small, or in other words, 
typical length scale of a system is large compared to $\ell_p$.

By varying the effective action (\ref{eq:R4}), equations of motion are obtained as
\begin{alignat}{3}
  E_{ij} &\equiv R_{ij} - \frac{1}{2} \eta_{ij} R + \gamma \Big\{
  - \frac{1}{2} \eta_{ij} \Big(t_8 t_8 R^4 - \frac{1}{4!} \epsilon_{11} \epsilon_{11} R^4 \Big) \notag
  \\
  &\quad
  + \frac{3}{2} R_{abci} X^{abc}{}_j - \frac{1}{2} R_{abcj}X^{abc}{}_i - 2 D_{(a} D_{b)} X^a{}_{ij}{}^b 
  \Big\} = 0. \label{eq:MEOM}
\end{alignat}
Here $D_a$ is a covariant derivative for local Lorentz indices and $X_{abcd}$ is defined as
\begin{alignat}{3}
  X_{abcd} &= \frac{1}{2} \big( X'_{[ab][cd]} + X'_{[cd][ab]} \big), \label{eq:X}
  \\
  X'_{abcd} &= 96 \big(
  R_{abcd} R_{efgh} R_{efgh} - 16 R_{abce} R_{dfgh} R_{efgh} + 2 R_{abef} R_{cdgh} R_{efgh} \notag
  \\
  &\qquad\,
  + 16 R_{aecg} R_{bfdh} R_{efgh} - 16 R_{abeg} R_{cfeh} R_{dfgh} - 16 R_{efag} R_{efch} R_{gbhd} \notag
  \\
  &\qquad\,
  + 8 R_{abef} R_{cegh} R_{dfgh} \big) \notag.
\end{alignat}
The details of the derivation of the eq.~(\ref{eq:MEOM}) can be found in ref.~\cite{Hyakutake:2013vwa}.
In the following, we construct solutions of the eq.~(\ref{eq:MEOM}) up to the linear order of $\gamma$.
If we wish to go beyond the linear order approximation, we should add more and more higher derivative terms
in general\cite{Green:2006gt}.
The linear order approximation is reliable when $\ell_s$ is small and $g_s$ is greater than 1. 
More precise argument can be found in ref.~\cite{Hyakutake:2014maa}.


\subsection{Smeared Quantum Black Hole}


The Schwarzschild black hole is easily up lifted into 11 dimensions
by adding extra 7 coordinates $y_m (m=4,\cdots,9)$ and $z$.
Here the coordinate $z$ is the 11th direction which is compactified on the circle with radius $R_{11}=\ell_s g_s$.
The metric is given by
\begin{alignat}{3}
  ds_{11}^2 &= - F dt^2 + F^{-1} dr^2 + r^2 d\Omega_2^2 + dy_m^2 + dz^2, \qquad
  F = 1 - \frac{r_\text{h}}{r}. \label{eq:11BS}
\end{alignat}
This is the smeared black hole, which is spread over 7 spatial directions.
This becomes a solution of the eq.~(\ref{eq:MEOM}) if and only if we set $\gamma=0$.

Let us construct a smeared quantum black hole. 
Since the metric (\ref{eq:11BS}) is not a solution of the eq.~(\ref{eq:MEOM}),
we modify the ansatz as follows.
\begin{alignat}{3}
  &ds_{11}^2 = - G_1^{-1} F_1 dt^2 + F_1^{-1} dr^2 + r^2 d\Omega_2^2 + G_2 ( dy_m^2 + dz^2 ), \label{eq:11BSq}
  \\
  &F_1(x) = 1 - \frac{1}{x} + \frac{\gamma}{r_\text{h}^6} f_1(x), \qquad
  G_i(x) = 1 + \frac{\gamma}{r_\text{h}^6} g_i(x), \quad (i=1,2). \notag
\end{alignat}
Information of the quantum corrections of the smeared black hole
is encoded in 2 functions, $f_1(x)$ and $g_i(x)$. 
Here we defined a dimensionless radial coordinate as $x = \frac{r}{r_\text{h}}$.
By inserting the ansatz into the equations of motion (\ref{eq:MEOM}), we obtain 4 nontrivial equations
which come from $E_{tt}$, $E_{rr}$, $E_{\theta\theta}$ and $E_{zz}$ components.
With the aid of Mathematica, these 4 differential equations are explicitly written as
\begin{alignat}{3}
  E_1 &= - 4 x^{11} f_1' - 4 x^{10} f_1 + 14 x^{11}(1 - x) g_2'' + 7 x^{10} (3 - 4x) g_2' 
  - 442368 x + 463104 = 0, \notag
  \\
  E_2 &= 4 x^{11} f_1' + 4 x^{10} f_1 + 4 x^{10} (1 - x) g_1' - 7 x^{10} (3 - 4 x) g_2' 
  + 55296 x - 76032 = 0, \notag
  \\
  E_3 &= 2 x^{12} f_1'' + 4 x^{11} f_1' + 2 x^{11} (1 - x) g_1'' - x^{10} (1 + 2 x) g_1' \notag
  \\
  &\quad\,
  - 14 x^{11} (1 - x) g_2'' +14 x^{11} g_2' - 248832 x + 283392 = 0, \label{eq:eomfg}
  \\
  E_4 &= 2 x^{12} f_1'' + 8 x^{11} f_1' + 4 x^{10} f_1 + 2 x^{11} (1 - x) g_1'' + x^{10} (1 - 4 x) g_1' \notag
  \\
  &\quad\,
  - 12 x^{11} (1 - x) g_2'' - 12 x^{10} (1 - 2 x) g_2' - 6912 = 0. \notag
\end{alignat}
Note that the equations of motion are approximated up to the linear order of $\gamma$,
since we know the M-theory effective action only up to this order.
After some calculations, which are explained in detail in the appendix \ref{sec:solfg}, 
the solution for the above equations is given by
\begin{alignat}{3}
  f_1(x) &= \frac{448}{x} \Big\{ c + \sum_{n=1}^9 \frac{1}{n x^n} + \frac{216}{7 x^8} - \frac{215}{7 x^9} 
  + (1 - c') \log \Big( 1 - \frac{1}{x} \Big) \Big\} , \notag
  \\
  g_1(x) &= 896 \Big[ \sum_{n=1}^9 \frac{n+1}{n x^n} + \frac{12}{x^9} 
  + (1 - c') \Big\{ \log \Big(1 - \frac{1}{x}\Big) - \frac{1}{x - 1} \Big\} \Big], \label{eq:f1g1g2}
  \\
  g_2(x) &= 256 \Big\{ \sum_{n=1}^9 \frac{1}{n x^n} + (1 - c') \log \Big( 1 - \frac{1}{x} \Big) \Big\}, \notag
\end{alignat}
where $c$ and $c'$ are integral constants.
It is possible to add integral constants in $g_1(x)$ and $g_2(x)$ in addition to $c$ and $c'$, but
these are absorbed by redefinitions of coordinates.
From the requirement that the curvature is not singular at $x=1$, we choose $c'=1$.
We will fix the remaining constant $c$ in the next subsection.
Notice that the above solution is reliable when $\ell_p \ll r_\text{h}$.
It is necessary to take into account more and more corrections for small $r_\text{h}$.


\subsection{Smeared Quantum Black 0-brane}


In this subsection, first we review
the way to construct a smeared black 0-brane solution, which carries momentum along the 11th $z$ direction.
This can be done by boosting the smeared black hole along it. 
The boost is executed on $(t,z)$-plane and the metric becomes 
\begin{alignat}{3}
  ds_{11}^2 &= - F (\cosh \beta dt + \sinh \beta dz)^2 + (\sinh \beta dt + \cosh \beta dz)^2 
  + F^{-1} dr^2 + r^2 d\Omega_2^2 + dy_m^2 \notag
  \\
  &= - H^{-1} F dt^2 + F^{-1} dr^2 + r^2 d\Omega_2^2 + dy_m^2 
  + H \Big( dz + (1 - H^{-1}) \frac{\cosh\beta}{\sinh \beta} dt \Big)^2. \label{eq:11BS-b}
\end{alignat}
Here $\beta$ is a boost parameter and $H$ is a harmonic function of
\begin{alignat}{3}
  H = 1 + \frac{r_\text{h} \sinh^2 \beta}{r}.
\end{alignat}
This geometry corresponds to a smeared M-wave solution in 11 dimensions.
For later use, we often rewrite these parameters as
\begin{alignat}{3}
  r_\text{h} = r_- \alpha, \qquad \sinh^2 \beta = \frac{1}{\alpha}.
\end{alignat}
The mass and the momentum of the smeared M-wave are expressed by these two parameters,
as we will see in next section.

When the solution (\ref{eq:11BS-b}) is reduced to 10 dimensions, 
the metric, the dilaton field and the R-R 1-form field are written as
\begin{alignat}{3}
  &ds_{10}^2 = - H^{-\frac{1}{2}} F dt^2 + H^{\frac{1}{2}} \big( F^{-1} dr^2 
  + r^2 d\Omega_2^2 + dy_m^2 \big), \label{eq:smearD0}
  \\
  &e^\phi = H^\frac{3}{4}, \qquad C^{(1)} = \sqrt{1+\alpha} \, (1 - H^{-1}) dt. \notag
\end{alignat}
This is the smeared black 0-brane solution in 10 dimensions.
Of course, if we take $r_- \to 0$ with $\frac{r_- \alpha}{r}$ fixed, 
the charge goes to zero and the above solution becomes the smeared black hole (\ref{eq:11BS}).
On the other hand, if we take $\alpha \to 0$ with $\frac{r_-}{r}$ fixed, the above solution becomes an extremal
black 0-brane smeared over $y_m$ directions.

Now we boost the smeared quantum black hole along 11th $z$ direction. 
The boost is executed on $(t,z)$-plane and the metric becomes
\begin{alignat}{3}
  ds_{11}^2 &= - G_1^{-1} F_1 (\cosh \beta dt \!+\! \sinh \beta dz)^2 
  \!+\! G_2 (\sinh \beta dt \!+\! \cosh \beta dz)^2 \!+\! F_1^{-1} dr^2 \!+\! r^2 d\Omega_2^2 \!+\! G_2 dy_m^2 \notag
  \\[0.1cm]
  &= - H_1^{-1} F_1 dt^2 \!+\! F_1^{-1} dr^2 + r^2 d\Omega_2^2 + G_2 dy_m^2 
  + H_2 \big( dz + \sqrt{1+\alpha} \big( 1 \!-\! H_3^{-1} \big) dt \big)^2. 
  \label{eq:met_bs}
\end{alignat}
Here $H_i \,(i=1,2,3)$ are functions of $x = \frac{r}{r_- \alpha}$, and by using 
$H(x) = 1 + \frac{1}{\alpha x}$, $f_1(x)$, $g_1(x)$ and $g_2(x)$, they are expressed as
\begin{alignat}{3}
  H_1 &= G_1 G_2^{-1} H_2 = 
  H + \frac{\gamma}{r_-^6 \alpha^6} \Big[ g_1 + \frac{1}{\alpha} 
  \Big\{ g_1 + \Big( 1 - \frac{1}{x} \Big) g_2 - f_1 \Big\} \Big], \notag
  \\
  H_2 &= G_2 + (G_2 - G_1^{-1} F_1) \sinh^2\beta = 
  H + \frac{\gamma}{r_-^6 \alpha^6} \Big[ g_2 + \frac{1}{\alpha} 
  \Big\{ \Big( 1 - \frac{1}{x} \Big) g_1 + g_2 - f_1 \Big\} \Big], \label{eq:Hs}
  \\
  H_3 &= G_2^{-1} H_2 =
  H + \frac{\gamma}{r_-^6 \alpha^7} \Big\{ \Big( 1 - \frac{1}{x} \Big) g_1 
  + \Big( 1 - \frac{1}{x} \Big) g_2 - f_1 \Big\}, \notag
\end{alignat}
up to the linear order of $\gamma$.
We call this geometry a smeared quantum M-wave solution in the M-theory.
The mass and the momentum along the 11th direction are expressed by two parameters $r_-$ and $\alpha$.

By reducing the 11th direction, we obtain a smeared quantum black 0-brane in the type IIA superstring theory.
The metric, the dilaton field and the R-R 1-form field are given by
\begin{alignat}{3}
  &ds_{10}^2 = - H_1^{-1} H_2^\frac{1}{2} F_1 dt^2 + H_2^{\frac{1}{2}} \big( F_1^{-1} dr^2 
  + r^2 d\theta^2 + r^2 \sin^2\theta d\phi^2 + G_2 dy_m^2 \big), \label{eq:smearQD0}
  \\
  &e^\phi = H_2^\frac{3}{4}, \qquad C^{(1)} = \sqrt{1+\alpha} \big( 1 - H_3^{-1} \big) dt. \notag
\end{alignat}
Note that the quantum corrections considered in the above are that of 1-loop amplitudes in the type IIA superstring theory.
There is a tree level correction coming from $\ell_s^6 R^4$ terms in the effective action of the type IIA superstring theory, but
it is not considered here because its form is not known so far.

Now we consider two limits of the above solution.
First, by taking $r_- \to 0$ with $\frac{r_- \alpha}{r}$ fixed, 
the charge goes to zero and the above solution corresponds to the smeared quantum black hole (\ref{eq:11BSq}).
Second, by taking $\alpha \to 0$ with $\frac{r_-}{r}$ and $\frac{\gamma}{r_-^6\alpha^6}$ fixed, 
the above solution becomes the extremal black 0-brane smeared over $y_m$ directions
if $g_1 + g_2 -f_1 = \mathcal{O}(x^{-2})$.
This is achieved when $c=\frac{32}{7}$ and $c'=1$ in the eq.~(\ref{eq:f1g1g2}).
Therefore $f_1(x)$, $g_1(x)$ and $g_2(x)$ are given by
\begin{alignat}{3}
  f_1(x) &= \frac{448}{x} \Big( \frac{32}{7} + \sum_{n=1}^9 \frac{1}{n x^n} + \frac{216}{7 x^8} - \frac{215}{7 x^9} \Big), \notag
  \\
  g_1(x) &= 896 \Big( \sum_{n=1}^9 \frac{n+1}{n x^n} + \frac{12}{x^9} \Big), \label{eq:f1g1g2-2}
  \\
  g_2(x) &= 256 \sum_{n=1}^9 \frac{1}{n x^n}. \notag
\end{alignat}
These functions capture quantum features of the smeared quantum black hole and the smeared quantum black 0-brane.
If the solution (\ref{eq:smearQD0}) is reduced to 4 dimensions, it becomes a charged quantum black hole in 4 dimensions.


\subsection{Near Horizon Limit}


Let us consider the near horizon limit of the smeared quantum black 0-brane (\ref{eq:smearQD0}).
The near horizon limit $r \to 0$ is defined so that physical quantities of the dual gauge theory become finite\cite{Maldacena:1997re}.
Thus we fix the energy scale $U$ and 't Hooft coupling $\lambda$, which are given by\cite{Itzhaki:1998dd}
\begin{alignat}{3}
  U = \frac{r}{\ell_s^2} ,\qquad \lambda = \frac{g_s N}{(2\pi)^2 \ell_s^3}.
\end{alignat}
$N$ is the number of the smeared D0-branes.
Note that the energy scale at the horizon $U_\text{h} = \frac{r_\text{h}}{\ell_s^2}$ is also fixed.
In terms of $\alpha$, $r_-$ and $\gamma$, the near horizon limit is defined as
\begin{alignat}{3}
  \alpha \to 0 \quad \text{with $x=\frac{r}{r_- \alpha}$, $\frac{r_- \alpha}{\ell_s^2}$ and $\frac{\gamma}{r_-^6\alpha^6}$ fixed.} 
  \label{eq:lim}
\end{alignat}
Now we are ready to take the near horizon limit of the eq.~(\ref{eq:smearQD0}). Then $H_i (i=1,2,3)$ are expressed as
\begin{alignat}{3}
  H_1 &= \frac{1}{\alpha} \Big[ \frac{1}{x}
  + \frac{\gamma}{r_-^6\alpha^6} \Big\{ g_1 + \Big( 1 - \frac{1}{x} \Big) g_2 - f_1 \Big\} \Big], \notag
  \\
  H_2 &= \frac{1}{\alpha} \Big[ \frac{1}{x}
  + \frac{\gamma}{r_-^6\alpha^6} \Big\{ \Big( 1 - \frac{1}{x} \Big) g_1 + g_2 - f_1 \Big\} \Big], \label{eq:HsNH}
  \\
  H_3 &= \frac{1}{\alpha} \Big[ \frac{1}{x}
  + \frac{\gamma}{r_-^6\alpha^6} \Big\{ \Big( 1 - \frac{1}{x} \Big) g_1 
  + \Big( 1 - \frac{1}{x} \Big) g_2 - f_1 \Big\} \Big], \notag
\end{alignat}
while $F_1(x)$ and $G_2(x)$ remain the same.
It is possible to check that the above becomes the solution of the eq.~(\ref{eq:MEOM}) directly,
although we do not show the details of the calculations.


\section{Thermodynamics of Smeared Quantum Black Objects} \label{sec:T-QBO}


In this section, we consider the thermodynamics of the black objects constructed in the previous section.
We see that the mass and the entropy of these objects are modified due to the quantum corrections.


\subsection{Thermodynamics of Smeared Quantum Black Hole}


Let us investigate the thermodynamics of the quantum black hole (\ref{eq:11BSq}) with the eq.~(\ref{eq:f1g1g2-2}).
The event horizon is located at $r_\text{horizon}=r_\text{h} - \frac{\gamma}{r_\text{h}^5} f_1(1)$
up to the linear order of $\gamma$, and the temperature $T$ is given by
\begin{alignat}{3}
  T &= \frac{1}{4\pi} G_1^{-\frac{1}{2}} \frac{d F_1}{dr} \Big|_{r_\text{horizon}} \notag
  \\
  &= \frac{1}{4\pi r_\text{h}} \Big( 1 + \gamma \frac{4f_1(1) + 2f'_1(1) - g_1(1)}{2r_\text{h}^6} \Big) \notag
  \\
  &= \frac{1}{4\pi r_\text{h}} \Big( 1 + \gamma \frac{1920}{r_\text{h}^6} \Big). \label{eq:Tmp}
\end{alignat}
By solving the above equation inversely, $r_\text{h}$ is expressed as
\begin{alignat}{3}
  r_\text{h} &= \frac{1}{4\pi T} \Big( 1 + \gamma \frac{4f_1(1) + 2f'_1(1) - g_1(1)}{2} (4\pi T)^6 \Big) \notag
  \\
  &= \frac{1}{4\pi T} \big( 1 + 1920 \gamma (4\pi T)^6 \big), \label{eq:r_h}
\end{alignat}
up to the linear order of $\gamma$.
By using the above relation, physical quantities can be expressed in terms of the temperature
up to the linear order of $\gamma$.
For instance, the location of the event horizon is evaluated as
\begin{alignat}{3}
  r_\text{horizon} &= \frac{1}{4\pi T} 
  \Big( 1 + \gamma \frac{2f_1(1) + 2f'_1(1) - g_1(1)}{2} (4\pi T)^6 \Big) \notag
  \\
  &= \frac{1}{4\pi T} \Big( 1 - \frac{65672}{45} \gamma (4\pi T)^6 \Big). \label{eq:r_horizon}
\end{alignat}
This reveals that the position of the event horizon slightly moves inward due to the quantum correction
for fixed temperature.

Since the solution (\ref{eq:11BSq}) is asymptotically flat, the mass is evaluated by ADM mass formula.
With the aid of Mathematica, the ADM mass $M$ is calculated as
\begin{alignat}{3}
  \frac{2\kappa_{10}^2}{4\pi V_6} M &= r_\text{h} \Big( 2 - \gamma \frac{2304}{r_\text{h}^6} \Big) \notag
  \\
  &= \frac{1}{2\pi T} \big( 1 + 768 \gamma (4\pi T)^6 \big). \label{eq:M_h}
\end{alignat}
Note that although the M-theory effective action (\ref{eq:R4}) contains the higher derivative terms,
the expression of the ADM mass formula is not affected\cite{Hyakutake:2014maa}.
On the other hand, the area law of the entropy is modified by the higher derivative corrections \cite{Wald:1993nt,Iyer:1994ys},
and the black hole entropy $S$ is given by
\begin{alignat}{3}
  \frac{2\kappa_{10}^2}{4\pi V_6} S &= 
  4\pi r^2 G_2^{\frac{7}{2}} \Big( 1 - 2 \gamma X_{0101} \Big) \Big|_\text{horizon} \notag
  \\
  &= 4\pi r_\text{h}^2 \Big( 1 - \gamma \frac{1920}{r_\text{h}^6} \Big) \notag
  \\
  &= \frac{1}{4\pi T^2} \big( 1 + 1920 \gamma (4\pi T)^6 \big). \label{eq:S_h}
\end{alignat}
It is easy to see that the first law of the black hole thermodynamics, $dM = T dS$, 
holds up to the linear order of $\gamma$.
Note that the specific heat is negative for the Schwarzschild black hole, but the quantum corrections
might make it positive at high temperature. Of course the above expression is reliable when $\gamma (4\pi T)^6 \ll 1$,
so for higher temperature we need to take into account higher order quantum corrections.


\subsection{Thermodynamics of Smeared Quantum Black 0-brane} \label{sec:T-SQD0}


Let us evaluate physical quantities of the smeared quantum black 0-brane (\ref{eq:smearQD0}) 
with the eq.~(\ref{eq:f1g1g2-2}) and the eq.~(\ref{eq:Hs}).
The event horizon is located at $r_\text{horizon} = r_\text{h} - \frac{\gamma}{r_\text{h}^5} f_1(1)$ 
as in the previous subsection. Remind that $r_\text{h} = r_- \alpha$.
The temperature is given by
\begin{alignat}{3}
  T &= \frac{1}{4\pi} H_1^{-1/2} \frac{dF_1}{dr} \Big|_{r_\text{horizon}} \notag
  \\
  &= \frac{1}{4\pi r_\text{h}} \sqrt{\frac{\alpha}{1+\alpha}} 
  \Big( 1 + \gamma \frac{1920}{r_\text{h}^6} \Big). \label{eq:Tsmear}
\end{alignat}
And the electric potential (or chemical potential) $\Phi$ is calculated as
\begin{alignat}{3}
  \Phi = C^{(1)}_t \Big|_{r_\text{horizon}} = \frac{1}{\sqrt{1+\alpha}}.
\end{alignat}
With the aid of Mathematica, the mass $M$ and the R-R charge $Q$ of the smeared black 0-brane are evaluated as
\begin{alignat}{3}
  M &= \frac{4\pi V_6}{2 \kappa_{10}^2} r_\text{h} 
  \Big( 2 + \frac{1}{\alpha} - \gamma \frac{2304}{r_\text{h}^6} \Big), \qquad
  Q = \frac{4\pi V_6}{2 \kappa_{10}^2} \frac{\sqrt{1 + \alpha}}{\alpha} r_\text{h}.
\end{alignat}
$V_6$ is the volume of the compactified 6 directions. 
As noticed before, formulae for the ADM mass and the charge are not affected by the quantum effect 
of the action (\ref{eq:R4}). The correction of the mass comes from the shift of the event horizon.
On the other hand, the entropy is modified by the action (\ref{eq:R4}). By using Wald's formula we obtain
\begin{alignat}{3}
  S = \frac{4\pi V_6}{2 \kappa_{10}^2} 4\pi r_\text{h}^2 \sqrt{\frac{1+\alpha}{\alpha}}
  \Big( 1 - \gamma \frac{1920}{r_\text{h}^6} \Big).
\end{alignat}
From these quantities, 
it is possible to see that the first law of the black hole thermodynamics, 
$\delta M = T \delta S + \Phi \delta Q$, is satisfied.
It is interesting to note that the R-R charge is not renormalized but the mass is. 
This may be true since we just take into account the quantum corrections of strings
which are neutral under the R-R charge.


\subsection{Near Horizon Limit}


Let us consider the thermodynamics of the near horizon geometry of the smeared black 0-brane (\ref{eq:smearQD0})
with the eq.~(\ref{eq:f1g1g2-2}) and the eq.~(\ref{eq:HsNH}).
First we introduce
\begin{alignat}{3}
  M_6 = \frac{V_6}{(2\pi\ell_s^2)^6}.
\end{alignat}
$M_6^{1/6}$ is an typical mass scale for the compactified 6 spatial directions, which is also fixed
in the near horizon limit (\ref{eq:lim}).
The R-R charge of the D0-branes are set to be
\begin{alignat}{3}
  Q = \frac{N}{\ell_s g_s},
\end{alignat}
where $N$ is a number of D0-branes. 
Then, in the near horizon limit, $\alpha$ and $\frac{\gamma}{r_\text{h}^6}$ behave like
\begin{alignat}{3}
  \alpha \;\to\; \frac{M_6 U_\text{h}}{2\pi^2 \lambda} \ell_s^4, \qquad
  \frac{\gamma}{r_\text{h}^6} \;\to\; \frac{\pi^6 \lambda^2}{2^7 3^2 N^2 U_\text{h}^6}.
\end{alignat}
The temperature (\ref{eq:Tsmear}) becomes
\begin{alignat}{3}
  T &= \frac{M_6^{1/2}}{4\sqrt{2}\pi^2 \lambda^{1/2} U_\text{h}^{1/2}}
  \Big( 1 + \frac{5 \pi^6 \lambda^2}{3 N^2 U_\text{h}^6} \Big), \label{eq:Tnear}
\end{alignat}
and the internal energy $E = M-Q$ and the entropy $S$ are expressed as
\begin{alignat}{3}
  \frac{E}{N^2} &= \frac{3M_6 U_\text{h}}{16\pi^4 \lambda^2} 
  \Big( 1 - \frac{4 \pi^6 \lambda^2}{3N^2 U_\text{h}^6} \Big), \label{eq:ESnear}
  \\
  \frac{S}{N^2} &= \frac{\sqrt{2} M_6^{1/2} U_\text{h}^{3/2}}{2\pi^2 \lambda^{3/2}} 
  \Big( 1 - \frac{5 \pi^6 \lambda^2}{3 N^2 U_\text{h}^6} \Big). \notag
\end{alignat}
In terms of the temperature, the above are expressed as
\begin{alignat}{3}
  \frac{E}{N^2} &= \frac{3M_6^2}{2(2\pi)^8 \lambda^3 T^2} 
  \Big( 1 + \frac{2 (32 \pi^5)^6 \lambda^8}{N^2 M_6^6} T^{12} \Big), \label{eq:ESnearT}
  \\
  \frac{S}{N^2} &= \frac{M_6^2}{(2\pi)^8 \lambda^3 T^3} 
  \Big( 1 + \frac{10 (32 \pi^5)^6 \lambda^8}{3N^2 M_6^6} T^{12} \Big). \notag
\end{alignat}
It is interesting to note that the specific heat will become positive at high temperature,
although we need to examine higher order corrections to make sure.
Notice that, in the case of a quantum black 0-brane, the quantum correction becomes important at low temperature\cite{Hyakutake:2013vwa}.
The situation is opposite for the smeared quantum black 0-brane.


\section{Radiation of String and D0-brane from Smeared Quantum Black 0-brane} \label{sec:rad}



\subsection{Radiation of String}


One of important phenomena on the black hole physics is the Hawking radiation.
As in the case of the Schwarzschild black hole, the non-extremal black $0$-brane 
is unstable and gradually emits the energy by thermal radiation.
In the superstring theory, this phenomena corresponds to the emission of strings from thermal D0-branes.
Although there are several ways to understand the Hawking radiation,
it is simple to deal with a tunnelling process of a particle through the event horizon. 
This can be done by calculating the imaginary part of the action\cite{Parikh:1999mf}.
Here we apply the Hamilton-Jacobi method developed in ref.~\cite{Angheben:2005rm} 
to the string action\footnote{For a review of the tunnelling method, see ref.~\cite{Majhi:2011yi} for example.}. 

We consider Nambu-Goto action as the string action.
The Lagrangian for the string in the background of (\ref{eq:smearQD0}) is given by
\begin{alignat}{3}
  \mathcal{L} &= - T_s \int d\phi \sqrt{- \det 
  \begin{pmatrix} 
    - H_1^{-1} H_2^\frac{1}{2} F_1 + H_2^{\frac{1}{2}} \big( F_1^{-1} \dot{r}^2 + r^2 \dot{\theta}^2 \big) && 0 \\ 
    0 && H_2^{\frac{1}{2}} r^2 \sin^2\theta
  \end{pmatrix}} \notag
  \\
  &= - 2\pi T_s H_1^{-\frac{1}{2}} H_2^{\frac{1}{2}} F_1^{\frac{1}{2}} r \sin\theta \sqrt{1 - H_1 F_1^{-2} \dot{r}^2
  - H_1 F_1^{-1} r^2 \dot{\theta}^2},
\end{alignat}
where $T_s = 1/(2\pi\ell_s^2)$ represents the tension of the string.
In the above we assumed that the string is extended homogeneously along $\phi$ direction and moving toward $r$ and $\theta$ directions.
The momentum conjugate to $r$ and $\theta$ are given by
\begin{alignat}{3}
  p_r &= \frac{\pa \mathcal{L}}{\pa \dot{r}}
  = 2\pi T_s H_1^{-\frac{1}{2}} H_2^{\frac{1}{2}} F_1^{\frac{1}{2}} r \sin\theta
  \frac{H_1 F_1^{-2} \dot{r}}{\sqrt{1 - H_1 F_1^{-2} \dot{r}^2 - H_1 F_1^{-1} r^2 \dot{\theta}^2 }}, \notag
  \\
  p_\theta &= \frac{\pa \mathcal{L}}{\pa \dot{\theta}}
  = 2\pi T_s H_1^{-\frac{1}{2}} H_2^{\frac{1}{2}} F_1^{\frac{1}{2}} r \sin\theta
  \frac{H_1 F_1^{-1} r^2 \dot{\theta}}{\sqrt{1 - H_1 F_1^{-2} \dot{r}^2 - H_1 F_1^{-1} r^2 \dot{\theta}^2 }},
\end{alignat}
and the Hamiltonian is written as
\begin{alignat}{3}
  \mathcal{H} &= \sqrt{ \big( 2\pi T_s r \sin\theta \big)^2 H_1^{-1} H_2 F_1
  + \frac{p_r^2}{H_1 F_1^{-2}} + \frac{p_\theta^2}{H_1 F_1^{-1} r^2} }.
\end{alignat}
Note that the first term in the square root corresponds to the rest mass of the string,
which becomes zero at the event horizon.

Now let us consider the Hamilton-Jacobi equation.
\begin{alignat}{3}
  - \frac{\pa \mathcal{S}}{\pa t} = \mathcal{H}(r, p_r, p_\theta), \qquad p_r = \frac{\pa \mathcal{S}}{\pa r}, \qquad 
  p_\theta = \frac{\pa \mathcal{S}}{\pa \theta}. \label{eq:HJstring}
\end{alignat}
$\mathcal{S}$ is the Hamilton's principal function. In order to solve this equation, we set
\begin{alignat}{3}
  \mathcal{S} = - \delta E \, t + W(r,\theta),
\end{alignat}
where $\delta E > 0$ is the energy taken by the string.
Since we are interested in the tunnelling process of the string through the event horizon, and for that purpose,
we only need to evaluate the imaginary part of $W(r,\theta)$.
By inserting the above ansatz into the Hamilton-Jacobi equation, we obtain
\begin{alignat}{3}
  \frac{\partial W}{\partial r} = H_1^{\frac{1}{2}} F_1^{-1}
  \sqrt{(\delta E)^2 - \big( 2\pi T_s r \sin\theta \big)^2 \frac{H_2 F_1}{H_1} - \frac{F_1 p_\theta^2}{H_1 r^2}},
\end{alignat}
for the outgoing string. Here we assume that the angular momentum $p_\theta$ is not divergent 
around the event horizon.
Thus around the horizon, by using $F_1 \sim 0$, the function of $W(r,\theta)$ is approximately solved as
\begin{alignat}{3}
  W(r,\theta) &\sim \delta E \int dr \, H_1^{\frac{1}{2}} F_1^{-1} 
  = \frac{\delta E}{2\pi T} \int \frac{d\rho}{\rho}.
\end{alignat}
Here we made a coordinate transformation $d\rho = H_2^{\frac{1}{4}} F_1^{-\frac{1}{2}} dr$, which
gives $H_1^{-1} H_2^{\frac{1}{2}} F_1 \sim \frac{1}{4} H_1^{-1} (\frac{dF_1}{dr})^2 \big|_{r_\text{horizon}} \rho^2
= 4\pi^2 T^2 \rho^2$.
This procedure is similar to the derivation of the Hawking temperature $T$ in the eq.~(\ref{eq:Tsmear}),
so we find it in the last equality.
The region of the integral is chosen from inside to outside the horizon.
The imaginary part of $W(r,\theta)$ gives the tunnelling probability of the string $P_s$, which is evaluated as
\begin{alignat}{3}
  P_s &= e^{-2 \text{Im} W} = e^{-\frac{\delta E}{T}} = e^{-|\delta S|}.
\end{alignat}
We used the first law of the black hole thermodynamics for $\delta Q = 0$.
$\delta S < 0$ is a difference of the entropy of the smeared quantum black 0-brane due to the radiation of the string.


\subsection{Radiation of D0-brane}


In this subsection we examine a test D0-brane moving around the smeared quantum black 0-brane.
First let us consider the potential energy for the test D0-brane,
which is moving only along the radial direction.
With this assumption, the Lagrangian for the D0-brane in the background of (\ref{eq:smearQD0}) becomes
\begin{alignat}{3}
  \mathcal{L} &= - T_0 e^{-\phi} \sqrt{- g_{\mu\nu} \dot{x}^\mu \dot{x}^\nu} - T_0 C^{(1)}_t \notag
  \\
  &= - T_0 e^{-\phi} H_1^{-\frac{1}{2}} H_2^{\frac{1}{4}} F_1^{\frac{1}{2}} \sqrt{1 - H_1 F_1^{-2} \dot{r}^2}
  - T_0 \sqrt{1+\alpha} (1-H_3^{-1}).
\end{alignat}
The momentum conjugate to $r$ is defined as
\begin{alignat}{3}
  p_r = \frac{\pa \mathcal{L}}{\pa \dot{r}}
  = T_0 H_1^{-\frac{1}{2}} H_2^{-\frac{1}{2}} F_1^{\frac{1}{2}} 
  \frac{H_1 F_1^{-2} \dot{r}}{\sqrt{1 - H_1 F_1^{-2} \dot{r}^2}}.
\end{alignat}
By using the above equation, $\dot{r}$ is written by $p_r$, 
and the Hamiltonian of the D0-brane is evaluated as
\begin{alignat}{3}
  \mathcal{H} &= H_1^{-\frac{1}{2}} H_2^{-\frac{1}{2}} F_1^{\frac{1}{2}} \sqrt{T_0^2 + H_2 F_1 p_r^2}
  + T_0 \sqrt{1+\alpha} (1-H_3^{-1}).
\end{alignat}
If the momentum is small enough, we can expand the above and read off the potential energy as
\begin{alignat}{3}
  V &= T_0 H_1^{-\frac{1}{2}} H_2^{-\frac{1}{2}} F_1^{\frac{1}{2}} + T_0 \sqrt{1+\alpha} (1-H_3^{-1}).
\end{alignat}
The first term corresponds to the attractive force by the gravity and the second term
does to the repulsive force due to the R-R background.
In the classical (or $1 \ll x$) and near horizon limits, the potential becomes
\begin{alignat}{3}
  V - T_0 &\sim T_0 \alpha x (\sqrt{F} - 1) + T_0 \frac{\alpha}{2} \notag
  \\
  &= - \frac{M_6 U_\text{h} N}{64\pi^4 \lambda^2} \frac{1}{x} .
\end{alignat}
In the above, the rest mass of the D0-brane is subtracted.
The quantum corrections give $\mathcal{O}(x^{-2})$ contribution.

Below let us consider the Hawking radiation of a charged particle, that is, the radiation of the D0-brane.
As in the previous subsection, we consider the Hamilton-Jacobi equation.
\begin{alignat}{3}
  - \frac{\pa \mathcal{S}}{\pa t} = \mathcal{H}(r, p_r), \qquad p_r = 
  \frac{\pa \mathcal{S}}{\pa r}. \label{eq:HJsmear}
\end{alignat}
In order to solve this equation, we set
\begin{alignat}{3}
  \mathcal{S} = - \delta E \, t + W(r),
\end{alignat}
where $\delta E > 0$ is the energy taken by the D0-brane.
we consider the outgoing D0-brane through the event horizon. 
By inserting the above ansatz into the Hamilton-Jacobi equation, 
the function $W(r)$ is solved around the horizon as
\begin{alignat}{3}
  W(r) &= \int dr \, H_1^{\frac{1}{2}} F_1^{-1} \sqrt{
  \big( \delta E - T_0 C^{(1)}_t \big)^2 - T_0^2 H_1^{-1} H_2^{-1} F_1 } 
  \sim \frac{\delta E - T_0 \Phi}{2\pi T} \int \frac{d\rho}{\rho},
\end{alignat}
where $\Phi$ is the electric potential.
Here, as in the previous subsection, we used $d\rho = H_2^{\frac{1}{4}} F_1^{-\frac{1}{2}} dr$
and $H_1^{-1} H_2^{\frac{1}{2}} F_1 \sim \frac{1}{4} H_1^{-1} (\frac{dF_1}{dr})^2 \big|_{r_\text{horizon}} \rho^2
= 4\pi^2 T^2 \rho^2$.
The Hawking temperature is given by the eq.~(\ref{eq:Tsmear}).
The region of the integral is chosen from inside to outside the horizon.
Finally WKB approximation of the tunneling probability through the even horizon is 
estimated by the imaginary part of $W(r)$, and we obtain
\begin{alignat}{3}
  P &= e^{-2\,\text{Im}\,W} = e^{- \frac{\delta E - T_0 \Phi}{T}} = e^{-|\delta S|}.
\end{alignat}
Note that $\delta Q = T_0$ for the radiation of the test D0-brane.
We used the first law of the black hole thermodynamics in the last equality.
$\delta S < 0$ is a difference of the entropy of the smeared quantum black 0-brane due to the radiation of the D0-brane.
For the near horizon limit, $\delta S \sim N$ and the tunnelling process is quite suppressed for large $N$.


\section{Conclusion and Discussion} \label{sec:con}


In this paper we examined quantum aspects of black objects in the type IIA superstring theory and the M-theory.
We explicitly solved the equations of motion which include leading quantum corrections in the M-theory,
and constructed the smeared quantum black hole and the smeared quantum black 0-brane.
These black objects can be reduced to 4 dimensions, and correspond to Schwarzschild and charged black holes
with quantum corrections. We discussed quantum nature of these black objects, such as the black hole thermodynamics
and the Hawking radiation.

In section \ref{sec:QBO}, we reviewed the quantum corrections in the M-theory and constructed the smeared quantum black hole solution
up to the linear order of $\gamma$.
The solution is characterised by 3 functions $F_1(x)$ and $G_i(x) \,(i=1,2)$, 
which are shown in the eq.~(\ref{eq:f1g1g2-2}).
The smeared quantum black 0-brane is also derived by boosting the smeared quantum black hole along 11th direction
and reducing its direction. It is characterized by 3 functions $H_i(x) \,(i=1,2,3)$, 
and the explicit forms are showed in the eq.~(\ref{eq:Hs}).
The near horizon limit of the smeared quantum black 0-brane is also considered. 
Note that there are several integral constants in $F_1(x)$ and $G_i(x) \, (i=1,2)$, 
but these are completely fixed by imposing appropriate conditions.

In section \ref{sec:T-QBO}, we derived the ADM mass and the entropy of the smeared quantum black hole, 
and confirmed the first law of the thermodynamics. The specific heat of the Schwarzschild black hole is negative,
but it might be possible to make it positive at high temperature due to the quantum corrections.
In order to confirm this phenomena, we need to take into account higher order corrections.
We also derived the mass, the R-R charge and the entropy of the smeared quantum black 0-brane,
and confirmed the first law of the thermodynamics. It is interesting to note that the R-R charge is not renormalized,
but the mass is. In the near horizon limit, again we argued the possibility that the quantum corrections make the specific heat positive.

The radiations of the string and the D0-brane out of the smeared quantum black 0-brane are discussed in section \ref{sec:rad}.
We calculated the tunnelling probability of the string and the D0-brane through the event horizon by employing the Hamilton-Jacobi method.
Although the derivations are different, both analyses show that the thermal D0-branes emit the string and the D0-brane 
with the same Hawking temperature. This shows that the smeared quantum black 0-brane is a thermal object, 
and the emission of a D-brane from D-branes is likely to occur. (See appendix \ref{sec:radp}.)

Since we have studied the quantum aspects of the smeared black 0-brane in the near horizon limit,
it is interesting to examine the same physics from the gauge theory for the D0-branes,
which is often called BFSS matrix model\cite{Banks:1996vh}.
In refs.~\cite{Hanada:2013rga,Hanada:2016zxj}, numerical studies on the gauge/gravity duality for the D0-branes are
tested including leading quantum corrections. And in ref.~\cite{Berkowitz:2016jlq}, the duality is confirmed by lattice simulation 
including leading quantum corrections with $\alpha'(=\ell_s^2)$ dependence. 
Thus it is possible to do the same analysis for the smeared quantum black 0-brane,
and will be interesting to clarify the quantum nature of 4 dimensional charged black hole from the viewpoint of the gauge theory.
It is also interesting to construct a configuration of the black hole from the gauge theory.
This will be done by using the fuzzy sphere solution of the BFSS matrix model\cite{Kabat:1997im,Aoki:2015uha}.
Further analysis on the relation between the black hole and the fuzzy sphere will be reported in near future.


\section*{Acknowledgement}


The author would like to thank Masanori Hanada for interesting discussions.
The seminar at KIPMU was also useful to complete this paper.
This work was partially supported by the Ministry of Education, Science, 
Sports and Culture, Grant-in-Aid for Young Scientists (B) 24740140, 2012.

\appendix


\section{Solution of Eq.~(\ref{eq:eomfg})} \label{sec:solfg}


In this appendix we show how to solve the equations of motion (\ref{eq:eomfg}) in detail.
First of all, we obtain an equation for $g_2(x)$ by taking the following linear combination.
\begin{alignat}{3}
  0 &= - \frac{1}{18x^{10}} \big( E_1 - E_2 - 2 E_3 + 2 E_4 \big) \notag
  \\
  &= x(x-1) g_2'' + (2 x-1) g_2' + \frac{2304}{x^{10}} \notag
  \\
  &= \Big\{ x(x-1) g_2' - \frac{256}{x^9} \Big\}'.
\end{alignat}
From this we obtain
\begin{alignat}{3}
  g_2' = \frac{256}{x^{10}(x-1)} - \frac{256c'}{x(x-1)}
  = 256 \Big\{ - \sum_{n=1}^9 \frac{1}{x^{n+1}} + \frac{1-c'}{x(x-1)} \Big\},
\end{alignat}
where $c'$ is an integral constant. Then $g_2$ is solved as
\begin{alignat}{3}
  g_2(x) = 256 \Big\{c'' + \sum_{n=1}^9 \frac{1}{n x^n} + (1 - c') \log \Big( 1 - \frac{1}{x} \Big) \Big\}.
\end{alignat}
Second, it is possible to obtain an equation for $g_1(x)$ as follows.
\begin{alignat}{3}
  0 &= \frac{1}{4x^{10}(1-x)} (E_1 + E_2) \notag
  \\
  &= \frac{7}{2} x g_2'' + g_1' + \frac{96768}{x^{10}} \notag
  \\
  &= \Big( g_1 + \frac{7}{2} x g_2' - \frac{7}{2} g_2 - \frac{10752}{x^9} \Big)'.
\end{alignat}
And $g_1(x)$ is solved as
\begin{alignat}{3}
  g_1(x) &= 896 \Big[ c''' + \sum_{n=1}^9 \frac{n+1}{n x^n} + \frac{12}{x^9}
  + (1 - c') \Big\{ \log \Big( 1 - \frac{1}{x} \Big) - \frac{1}{x-1} \Big\} \Big].
\end{alignat}
Finally we examine $E_2 = 0$.
\begin{alignat}{3}
  0 &= \frac{E_2}{4x^{10}} \notag
  \\
  &= x f_1' + f_1 + (1-x) g_1' + \Big(7 x - \frac{21}{4} \Big) g_2' 
  - \frac{19008}{x^{10}} + \frac{13824}{x^9} \notag
  \\
  &= \Big\{ x f_1 + (1-x) g_1 + \Big(\frac{7}{2} x - \frac{21}{4} \Big) g_2 \Big\}' 
  + g_1 + \frac{7}{2} x g_2' - \frac{7}{2} g_2 - \frac{19008}{x^{10}} + \frac{13824}{x^9} \notag
  \\
  &= \Big\{ x f_1 + (1-x) g_1 + \frac{7}{2} \Big( x - \frac{3}{2} \Big) g_2 + 192 \Big( \frac{11}{x^9} - \frac{16}{x^8} \Big)
  + 896 (c'''-c'') x \Big\}' .
\end{alignat}
From this, we obtain
\begin{alignat}{3}
  f_1(x) &= \frac{448}{x} \Big\{ c + \sum_{n=1}^9 \frac{1}{n x^n} + \frac{216}{7 x^8} - \frac{215}{7 x^9} 
  + (1 - c') \log \Big( 1 - \frac{1}{x} \Big) \Big\}. 
\end{alignat}
Remaining equation is automatically solved.
As discussed in section \ref{sec:QBO}, we choose the constants as $c=\frac{32}{7}$, $c'=1$ and $c''=c'''=0$.


\section{Emission of D$p$-brane from Classical Black $p$-brane} \label{sec:radp}


Let us consider the emission of a single D$p$-brane from non-extremal black 
$p$-brane by using the Hamilton-Jacobi method.
In the type IIA supergravity, the classical black $p$-brane is expressed as
\begin{alignat}{3}
  &ds^2 = H^{-\frac{1}{2}} \big( - f dt^2 + dx_i^2 \big)
  + H^{\frac{1}{2}} \big( f^{-1} dr^2 + r^2 d\Omega_{8-p}^2 \big), \label{eq:blackp}
  \\[0.2cm]
  &e^\phi = H^{\frac{3-p}{4}}, \qquad
  C^{(p+1)} = \Big(\frac{r_+}{r_-}\Big)^\frac{7-p}{2} (1-H^{-1}) \, 
  dt \wedge dx^1 \wedge \cdots dx^p, \notag
  \\
  &H = 1 + \frac{r_-^{7-p}}{r^{7-p}}, \qquad 
  f = 1 - \frac{r_+^{7-p} - r_-^{7-p}}{r^{7-p}}. \notag
\end{alignat}
Here $t=x^0$ is the temporal direction and $x_i \, (i=1,\cdots,p)$ are spatial directions along the black $p$-brane.
There are two parameters $r_\pm$, which are related to ADM mass and the R-R charge of the classical black $p$-brane.
The event horizon is located at $r_\text{H} = (r_+^{7-p} - r_-^{7-p})^{1/(7-p)}$.

First, let us derive the Hamiltonian for the D$p$-brane. The action of the D$p$-brane is given by
\begin{alignat}{3}
  \mathcal{S}_p = - T_p \int d^{p+1} x \, e^{-\phi} 
  \sqrt{-\det \big( g_{\mu\nu} \pa_a X^\mu \pa_b X^\nu \big)} 
  - T_p \int dx^{p+1} C_{0 \cdots p}, \label{eq:Dpact}
\end{alignat}
where $T_p^{-1} = (2\pi)^p\ell_s^{p+1} g_s$ and $a,b=0,\cdots,p$.
Now we consider the black $p$-brane background (\ref{eq:blackp}) and take the static gauge $X^a = x^a$.
Below we examine the situation where the D$p$-brane moves along the radial direction only, 
that is, the position of the D$p$-brane is given by $r(t)$.
Then the above action is written as
\begin{alignat}{3}
  \mathcal{S}_p = - q_p \int dt \, 
  \Big\{ H^{-1} f^{\frac{1}{2}} \sqrt{1 - H f^{-2} \dot{r}^2} 
  + \Big(\frac{r_+}{r_-}\Big)^\frac{7-p}{2} (1 - H^{-1}) \Big\},
\end{alignat}
where $q_p = T_p V_p$ and $V_p$ is the volume of the D$p$-brane.
As usual, the momentum conjugate to $r$ is defined as
\begin{alignat}{3}
  p_r = \frac{\pa \mathcal{L}_p}{\pa \dot{r}}
  = q_p H^{-1} f^{\frac{1}{2}} \frac{H f^{-2} \dot{r}}{\sqrt{1 - H f^{-2} \dot{r}^2}}.
\end{alignat}
By using the above equation, $\dot{r}$ is written by $p_r$, 
and the Hamiltonian of the D$p$-brane is evaluated as
\begin{alignat}{3}
  \mathcal{H}_p(r,p_r) &= H^{-1} f^{\frac{1}{2}} \sqrt{q_p^2 + H f p_r^2}
  + q_p \Big(\frac{r_+}{r_-}\Big)^\frac{7-p}{2} (1 - H^{-1}) .
\end{alignat}
Note that R-R force is repulsive since black $p$-brane and the D$p$-brane
carry the positive R-R charge.

Second, let us consider the Hamilton-Jacobi equation.
\begin{alignat}{3}
  - \frac{\pa \mathcal{S}_p}{\pa t} = \mathcal{H}_p(r, p_r), \qquad p_r = 
  \frac{\pa \mathcal{S}_p}{\pa r}. \label{eq:HJ}
\end{alignat}
In order to solve this equation, we set
\begin{alignat}{3}
  \mathcal{S}_p = - \delta E \, t + W(r).
\end{alignat}
This is the ansatz for outgoing mode through the event horizon. 
By inserting the above ansatz into the Hamilton-Jacobi equation, 
the function $W(r)$ is solved as
\begin{alignat}{3}
  W(r) &= \int dr \, H^{\frac{1}{2}} f^{-1} \sqrt{
  \Big\{\delta E - q_p \Big(\frac{r_+}{r_-}\Big)^\frac{7-p}{2} (1 - H^{-1})\Big\}^2
  - q_p^2 H^{-2} f } \notag
  \\
  &\sim \frac{1}{2\pi T_p} \int \frac{d\rho}{\rho} \, \sqrt{
  \Big\{\delta E - q_p \Big(\frac{r_+}{r_-}\Big)^\frac{7-p}{2} (1 - H^{-1}) \Big\}^2
  - q_p^2 H^{-2} f }. \notag
\end{alignat}
Here we introduced $d\rho = H^{\frac{1}{4}} f^{-\frac{1}{2}} dr$
and used $H^{-\frac{1}{2}} f \sim \frac{1}{4} H^{-1} (\frac{df}{dr})^2 \big|_{r_\text{H}} \rho^2
= 4\pi^2 T_p^2 \rho^2$.
$T_p$ represents the Hawking temperature.
The region of the integral is chosen from inside to outside the horizon.

Finally WKB approximation of the tunneling probability through the event horizon is 
estimated by evaluating the imaginary part of $W(r)$, and we obtain
\begin{alignat}{3}
  P &= e^{-2\,\text{Im}\,W} = e^{- \frac{\delta E - q_p \Phi_p}{T_p}} = e^{-|\delta S|}.
\end{alignat}
Here we defined the electric potential between the spatial infinity and the horizon as
$\Phi_p = C^{(p+1)}_{01\cdots p} \big|_{r_\text{H}} = \big(\frac{r_-}{r_+}\big)^\frac{7-p}{2}$.
We used the first law of the black $p$-brane thermodynamics in the last equality.


\end{document}